# ROOTING CAPACITY OF HARDWOOD CUTTINGS OF SOME FRUIT TREES IN RELATION TO CUTTING PATTERN

Aram Akram Mohammed, Rasul Rafiq Aziz, Faraydwn Karim Ahmad, Ibrahim Maaroof Noori and Tariq Abubakr Ahmad
Dept. of Horticulture, College of Agricultural Sciences Engineering, University of Sulaimani,
Kurdistan Region-Iraq

**ABSTRACT**

study two cut patterns in hardwood cuttings of (*Cydonia oblonga*), (*Punica granatum*) and (*Ficus carica*). The cuttings have been cut either straight with different internode stub lengths [0 (just onto the basal node as control), 0.5, 1.0, 2.0 or 3.0 cm below the basal node], or slant with 45° angle for each length mentioned above (except the first length (0 cm). Effect of the basal cut directions on rooting percentage and other shoot and root characteristics were not significantly different, while the effect of slant cut pattern on one-side rooting at the basal margin observed in some quince cuttings but it was rarely observed in pomegranate and fig cuttings. Quince cuttings gave no different rooting percentage and other shoot and root characteristics significantly with different internode stub lengths. While, internode stub 1 and 2 cm in pomegranate cuttings, and 0 cm in fig cuttings gave the best rooting percentages 44.44% and 100%, respectively. Also, interaction effects of the two factors on rooting percentage and other shoot and root characteristics were just significantly different in pomegranate and fig cuttings. The best rooting capacity achieved in pomegranate cuttings (49.99%) in those were cut straightly at the base with 1 and 2 cm basal internode stub lengths, and fig cuttings straightly cut at the base with 0 and 1 cm basal internode stub lengths gave the highest rooting capacity (100%).

## INTRODUCTION

Techniques in taking stem cuttings sometimes include some characteristics related to stem cutting themselves, for example length of cuttings, diameter, position of cuttings from shoots, age of the shoot from which cuttings are taken, retain or excise of leaves, type of stem cuttings, and wounding of cutting bases, all techniques are effective to form root in cuttings. In spite of these, cut patterns of stem cuttings is another characteristic may influence root initiation of the cuttings (Bruck and Paolillo, 1984 and Tchoundjeu and Leakey, 1996). Patterns of cut during taking cuttings include slant cut at basal or distal end of cuttings and length of remaining internode stub below the basal node, particularly in hardwood cuttings. However, recommendations to use the two patterns are very variable in researches and literatures of plant propagation. Bray *et al*. (2003) took hardwood cuttings of Arkansas blackberries with approximately 1.5 cm below the basal node, while Sharma and Aier (1989) took hardwood cuttings of Japanese plum with 0.5 cm below bud of the basal end. Also, Hartmann *et al*. (2011) demonstrated that basal cut is usually just below the basal node of the cuttings. In horticulture, most commonly, cuttings are cut just below a node because nodes on the stem are sites of adventitious root formation, besides adventitious root may produce from callus grow at the wounding sites as well (Rudall, 2007).

On the other hand, slant cut on one end is usually used in cuttings to recognize the upper and lower end of the cuttings, particularly in leafless hardwood cuttings. Whereas, some researchers prefer to give slant cut to the base of cuttings and others straight. Kumar (2016) referred that slant cut of basal cuttings give opportunity to expose cuttings for hormone treatment and root development with a larger wounding area. Inversely, Jaenicke and Beniest (2002) instructed to avoid slant cut at the base of cuttings, because this may led to produce roots in one-side. For these reasons, the objective of this

aram.hamarashed@univsul.edu.iq

study is to investigate and show whether cut length of remaining internode stub below the basal node and slant cut at one end of hardwood cuttings may have effect on rooting and other traits in pomegranate, fig and quince hardwood cuttings.

## MATERIALS AND METHODS

This research was carried out in the College of Agricultural Engineering Sciences, University of Sulaimani to study two cut patterns in hardwood cuttings of quince (*Cydonia oblonga*), pomegranate (*Punica granatum*) and fig (*Ficus carica*). The cuttings were collected from local cultivars of the three fruit types which grown in Kurdistan region, the quince cultivar known as (Baghdadi), the pomegranate (Salakhani) and the fig (Shahanjir). All hardwood cuttings were taken from basal parts of one-year-old shoots of the three fruit types on February 13, 2019.

The cuttings have been cut either straight with different internode stub lengths [0 (just onto the basal node), 0.5, 1.0, 2.0 or 3.0 cm below the basal node], or slant with 45° angle for each length mentioned above (except the first length (0 cm), due to the inability of making the cutting angle). The length and diameter of the cuttings were 20 cm and 0.7-1.1 mm, respectively. The bases of the cuttings were treated with 1:8 captan 50% fungicide in talc powder, and then planted in sand medium put in black polyethylene bags, in tunnel.

The study was laid down in RCBD with three replications, the cuttings were planted and covered with plastic sheets until April 1, 2019 to protect from further lowering of temperature. The cuttings were planted in black polyethylene bags, and in each bag, 6 cuttings were planted. For all fruit types, treatments, and replications, 81 plastic bags (486 cuttings) were used. After 104 days (on May 28, 2019) the cuttings uprooted and the parameters of rooting percentage, root length, root number, shoot length, shoot diameter were measured. After that, the cuttings transplanted in mixed medium of soil and compost in plastic bags in a greenhouse with a temperature range of (19-35 °C) for calculation survival percentage after transplanting, and survival percentage after transplanting was calculated after 26 days of the transplanting.

The data analyzed by computer XLSTAT software and Duncun's multiple range test (P≤0.05) was used for comparison of means. The comparison and interaction were done between the two cut patterns for each fruit type separately.

## RESULT AND DISCUSSION

Effects of basal cut direction on rooting percentage of hardwood cuttings of the quince, pomegranate and fig are shown in table (1). It was demonstrated that the effects of straight and slant cut directions of the bases of hardwood cuttings were not significant. Also, straight and slant cut direction of the bases of the cuttings did not have significant effect on other studied parameters, except in survival percentage in fig cuttings, the highest (84%) cuttings survived after transplanting in cuttings that cut in straight pattern at the basal end, and the lowest (77.49%) in cuttings cut in slant pattern at the basal end, whereas the cuttings of quince and pomegranate (100%) survived either they were cut in a straight or slant pattern. However, in some cuttings, especially in quince cuttings, it was observed that the cuttings taken in slant pattern at basal end produced root at one-side, but it was rarely observed in pomegranate and fig cuttings. Similarly, in stem cuttings of *Khaya ivorensis*, it was observed that cuttings obliquely cut at the base were rooted in one-side, but it was not effective on rooting percentage and root number (Tchoundjeu and Leakey, 1996). Besides, Soylu and Ertürk (1997) discussed that rooting hardwood cuttings of filbert were not significantly affected by basal cut types.

**Table (1): Effect of basal cut direction of hardwood cuttings of quince, pomegranate and fig on root and shoot characteristics.**

| Fruit type | Basal cut direction | Rooting % | Root number | Root length (cm) | Shoot length (cm) | Shoot diameter (mm) | Survival % |
|---|---|---|---|---|---|---|---|
| Quince | Straight | 84.44 a | 13.82 a | 13.23 a | 16.44 a | 2.42 a | 100 a |
|  | Slant | 91.66 a | 13.72 a | 13.03 a | 16.36 a | 2.39 a | 100 a |
| Pomegranate | Straight | 37.77 a | 8.12 a | 2.12 a | 8.72 a | 2.08 a | 100 a |
|  | Slant | 33.33 a | 6.77 a | 1.67 a | 7.03 a | 1.91 a | 100 a |
| Fig | Straight | 93.33 a | 24.01 a | 10.67 a | 6.99 a | 4.39 a | 84 a |
|  | Slant | 86.10 a | 22.90 a | 10.69 a | 6.45 a | 5.22 a | 77.49 b |

\* The values in each column with the same letter do not differ significantly ($P \leq 0.05$) according to Duncan's Multiple Range Test.

The results in table (2) explain that quince cuttings gave no different rooting percentage significantly with different internode stub lengths below the basal node. While, rooting percentage in pomegranate cuttings with different basal internode stub lengths were significantly different, the highest (44.44%) rooting was obtained in cuttings with 1 and 2 cm basal internode stub lengths, and the lowest (22.21%) rooting was obtained in cuttings with 0.5 cm basal internode stub length. Additionally, effects of different basal internode stub lengths on fig cuttings showed that cuttings with 0 cm basal internode stub length (straight cut just below the basal node without any internode stub) gave (100%) rooting, but cuttings with 3 cm basal internode stub length gave minimum (83.33%) rooting. In this regard, Lovell and White (1986) mentioned that root primordia in cuttings of woody plants may produce at internodes or nodes and the length of internode below the basal node is probably affected the rate of root occurrence at these locations, and also they referred that roots occasionally emerge very close to nodes at basal end. In the present study, fig cuttings confirmed that rooting was the best just at basal node, this may be belonged to the nodes which are places of higher auxin accumulation, which long ago known, it has a great role in adventitious root formation in cuttings. Bruck and Paolillo (1984) found that production of a high rooting rate at locations near the basal nodes may be due to nodes which are sites of high auxin concentration. On the contrary, the rooting rate in quince and pomegranate cuttings did not support the mentioned findings. On the other hand, the differences in rooting rate among the different fruit species with different basal internode stub lengths in this study may be due to their differences in the structures between nodes and internodes from which root arise. Geiss *et al.* (2009) described that roots in cuttings arise from a group of cells, called root initials, occurred in different tissues of the stem according to the species. Brutsch *et al.* (1977) noted that roots greatly formed from callus produced in basal cut or intermodal tissues rather than leaf traces. Inversely, in *Griselinia littoralis* and *G. lucida* cuttings, root initiation occurred a few millimeters above the basal cut, and leaf traces was the place of the highest activity of cambium (White and Lovell, 1984).

As it is shown in the same table, root number was significantly different in cuttings of the three fruit trees with different basal internode stub lengths. Quince cuttings with 0 cm, pomegranate and fig cuttings with 2 cm basal internode stub length gave the highest (18.95, 12.55 and 27.58 roots, respectively), whereas cuttings with 3 cm basal internode stub length gave the lowest (9.47 and 2.96) root number for quince and pomegranate, respectively, however fig cuttings showed the lowest (17.43) root number in cuttings with 1 cm basal internode stub length. Cuttings of quince and pomegranate showed no different root length statistically, but the fig cuttings showed significant differences. The longest (13.04 cm) fig root observed in cuttings with 2 cm and the shortest (8.01 cm) root was noticed in cuttings with 1 cm basal internode stub length. Shoot length and diameter were not significantly different for cuttings of the three fruits. Survival percentage after transplanting was just different in fig cuttings, fig cuttings with 0 cm basal internode stub length (100%) survived, cuttings survival with 1 cm basal internode stub length were minimum (68.33%), this may be due to 1 cm length of basal internode stub gave the lowest root number and length, hence little water and nutrients may be taken by these cuttings. Whereas, survival percentage was 100% for quince and pomegranate cuttings for all basal internode stub lengths.

**Table (2):** Effect of length of remaining internode stub from the basal node of hardwood cuttings of quince, pomegranate and fig on root and shoot characteristics.

| Fruit type | Basal stub length (cm) | Rooting % | Root number | Root length (cm) | Shoot length (cm) | Shoot diameter (mm) | Survival % |
|---|---|---|---|---|---|---|---|
| Quince | 0.0 | 77.77 a | 18.95 a | 12.11 a | 17.16 a | 2.61 a | 100 a |
| | 0.5 | 88.88 a | 13.96 b | 13.63 a | 17.66 a | 2.44 a | 100 a |
| | 1.0 | 88.88 a | 14.36b | 13.86 a | 17.30 a | 2.33 a | 100 a |
| | 2.0 | 83.33 a | 14.7 b | 13.05 a | 14.61 a | 2.44 a | 100 a |
| | 3.0 | 94.44 a | 9.47 c | 12.56 a | 15.69 a | 2.32 a | 100 a |
| Pomegranate | 0.0 | 33.33 ab | 7.72 ab | 1.22 a | 7.38 a | 2.23 a | 100 a |
| | 0.5 | 22.21 b | 7.33 ab | 2.25 a | 10.33 a | 2.12 a | 100 a |
| | 1.0 | 44.44 a | 7.16 ab | 2.02 a | 6.72 a | 1.91 a | 100 a |
| | 2.0 | 44.44 a | 12.55 a | 2.53 a | 8.70 a | 2.20 a | 100 a |
| | 3.0 | 33.33 ab | 2.96 b | 1.24 a | 6.41 a | 1.69 a | 100 a |
| Fig | 0.0 | 100 a | 21.11 ab | 8.44 ab | 6.35 a | 4.52 a | 100 a |
| | 0.5 | 94.44 ab | 26.05 a | 11.33 ab | 7.46 a | 4.42 a | 71.66 c |
| | 1.0 | 86.10 ab | 17.43 b | 8.01 b | 6.11 a | 5.79 a | 68.33 c |
| | 2.0 | 91.66 ab | 27.58 a | 13.04 a | 6.57 a | 4.24 a | 80 b |
| | 3.0 | 83.33b | 24.22 ab | 11.47 ab | 7.07 a | 2.70 a | 95 a |

**\*** The values in each column with the same letter do not differ significantly ($P \leq 0.05$) according to Duncan's Multiple Range Test.

Interaction effects of basal cut direction and length of internode below the basal node on rooting percentage in quince cuttings as shown in table (3) was not significant, the reason may belong to most roots in quince cuttings formed at the upper part (at the nodes) of the cuttings through the medium rather than at the basal end or basal margin for all basal internode stub lengths with the both basal cut directions. On the contrary, interaction of the two factors was significant on rooting percentage in pomegranate and fig cuttings. The best (49.99%) rooting was achieved in pomegranate cuttings taken straightly at the base with 1 and 2 cm basal internode stub lengths, while pomegranate cuttings that cut in slant pattern at the base with 0.5 cm basal internode stub length gave the lowest (16.66%) rooting. Interaction effects of the two factors on fig rooting percentage showed that fig cuttings taken straightly at the base with 0 and 1 cm basal internode stub lengths gave the highest (100%) rooting, while cuttings that cut in slant pattern at the base with 1 cm basal internode stub length gave the lowest (72.21%) rooting.

Also, effect of interaction of the two factors on root number in table (3) demonstrated that quince cuttings taken straightly with 0 cm (without internode stub below basal node) gave the maximum (18.95) root number, and cuttings that cut in slant pattern at the base with 3 cm basal internode stub length gave the lowest (9.24) root number. Straight cut pattern at the bases of cuttings with 2 cm basal internode stub length produced the most root number (13.88 and 29.05) in pomegranate and fig cuttings, respectively. Effect of interaction of the two factors on root length was just different in fig cuttings. The longest (15.38 cm) root achieved in fig cuttings that cut in straight pattern with 2 cm basal internode stub length, and the shortest (4.78 cm) in cuttings that cut in slant pattern with 1 cm basal internode stub lengths.

**Table (3): Effect of the interaction between basal cut direction and length of remaining internode stubs below the basal node of hardwood cuttings of quince, pomegranate and fig on root and shoot characteristics.**

| Basal cut direction | Basal internode stub length (cm) | Rooting % | Root number | Root length (cm) | Shoot length (cm) | Shoot diameter (mm) | Survival % |
|---|---|---|---|---|---|---|---|
| | | | | Quince | | | |
| Straight | 0.0 | 77.77 a | 18.95 a | 12.11 a | 17.16 ab | 2.61 a | 100 a |
| | 0.5 | 83.33 a | 12.86 bcd | 14.18 a | 19.49 a | 2.70 a | 100 a |
| | 1.0 | 88.88 a | 14.92 abc | 13.82 a | 17.18 a | 2.40 a | 100 a |
| | 2.0 | 77.77 a | 12.64 bcd | 14.15 a | 11.82 b | 2.28 a | 100 a |
| | 3.0 | 94.44 a | 9.71 cd | 11.92 a | 15.94 ab | 2.13 a | 100 a |
| Slant | 0.5 | 94.44 a | 15.05 abc | 13.07 a | 15.82 ab | 2.18 a | 100 a |
| | 1.0 | 88.888 a | 13.80 a-d | 13.90 a | 16.78 ab | 2.26 a | 100 a |
| | 2.0 | 88.88 a | 16.78 ab | 11.95 a | 17.41 ab | 2.61 a | 100 a |
| | 3.0 | 94.44 a | 9.24 d | 13.21 a | 15.45 ab | 2.52 a | 100 a |
| | | | | Pomegranate | | | |
| Straight | 0.0 | 33.33 ab | 7.72 ab | 1.22 a | 7.38 ab | 2.23 a | 100 a |
| | 0.5 | 27.77 ab | 10 ab | 3 a | 12.66 a | 1.96 a | 100 a |
| | 1.0 | 49.99 a | 8.48 ab | 2.63 a | 6.69 ab | 2.01 a | 100 a |
| | 2.0 | 49.99 a | 13.88 a | 2.34 a | 9.77 ab | 2.12 a | 100 a |
| | 3.0 | 27.77 ab | 3.22 b | 1.44 a | 7.10 ab | 2.10 a | 100 a |
| Slant | 0.5 | 16.66 b | 4.66 ab | 1.5 a | 8 ab | 2.28 a | 100 a |
| | 1.0 | 38.88 ab | 5.83 ab | 1.41 a | 6.75 ab | 1.81 a | 100 a |
| | 2.0 | 38.88 ab | 11.22 ab | 2.72 a | 7.64 ab | 2.27 a | 100 a |
| | 3.0 | 38.88 ab | 2.70 b | 1.05 a | 5.73 b | 1.29 a | 100 a |
| | | | | Fig | | | |
| Straight | 0.0 | 100 a | 21.11 a | 8.44 bc | 6.35 a | 4.52 a | 100 a |
| | 0.5 | 94.44 a | 26.58 a | 11.33 ab | 7.25 a | 4.43 a | 70 c |
| | 1.0 | 100 a | 26.05 a | 11.24 ab | 8.45 a | 4.99 a | 83.33 b |
| | 2.0 | 94.44 a | 29.05 a | 15.38 a | 6.38 a | 3.77 a | 100 a |
| | 3.0 | 83.33 ab | 20.22 a | 10.70 ab | 6.53 a | 4.25 a | 90 ab |
| Slant | 0.5 | 94.44 a | 25.53 a | 11.33 ab | 7.66 a | 4.41 a | 66.66 c |
| | 1.0 | 72.21 b | 8.82 b | 4.78 c | 3.76 b | 6.59 a | 60 c |
| | 2.0 | 88.88 ab | 26.10 a | 11.28 ab | 6.76 a | 4.72 a | 60 c |
| | 3.0 | 83.33 ab | 28.22 a | 11.66 ab | 7.61 a | 5.16 a | 100 a |

\* The values in each column with the same letter do not differ significantly ($P \leq 0.05$) according to Duncan's Multiple Range Test.

Quince and pomegranate cuttings (Table 3) which cut straightly at the base with 0.5 cm basal internode stub length gave the longest (19.49 and 12.66 cm, respectively) shoot. Quince cuttings which cut straightly at the base with 2 cm basal internode stub length gave the shortest (11.82 cm) shoot, and cuttings of pomegranate cut in slant pattern with 3 cm basal internode stub length gave the shortest (5.73 cm) shoot. Additionally, fig cuttings which cut straightly at the base with 1 cm basal internode stub length gave longest (8.45 cm) shoot, but those cut in slant pattern with the same length internode gave the shortest (3.76 cm) shoot. The same table showed that effect of interaction of the two factors was not significant on shoot diameter of the three fruit cuttings. While, Quince and pomegranate cuttings (100%) survived after transplanting indifferently at interaction of the two factors, but fig cuttings (100%) survived just in cuttings straightly cut at the base with 0 and 2 cm basal internode stub length, and those cut in slant pattern with 3 cm basal internode stub length.

## CONCLUSION

The results revealed that rooting percentage and the studied parameters of root and shoot characteristics in cuttings of three fruits were independent form basal cut direction patterns either straightly cut or cut in slant pattern. However, cuttings of the three fruits showed different results with different internode stub lengths below the basal node. According to the fruit type, quince cuttings did not affect by different basal internode stub lengths statistically, but pomegranate cuttings were the best with 1 and 2 cm internode stub lengths, and fig cuttings with 0 cm basal internode stub length. Generally, interaction effects of the two factors produced different results of the studied parameters in the cuttings of the three fruits, except in shoot diameter and survival percentage was just different in transplanted fig cuttings. Effects of the two cut patterns were different according to fruit types. However, using the cut patterns with cuttings of other fruit types may be helpful for further explanation the effect of cut patterns on root initiation in cuttings.